# Elastic polymer networks of exceptional strength by deconcentrating tension

Yu Zhou[†], Chenghai Li[†], Zheqi Chen[†], Xianyang Bao, Zhigang Suo*

John A. Paulson School of Engineering and Applied Sciences, Harvard University, MA 02138
* To whom correspondence should be addressed. Email: suo@seas.harvard.edu
[†] These authors contributed equally to this work.

**Abstract**
The strength of a polymer network is orders of magnitude below that of a polymer chain, because the network concentrates high tension in a small fraction of polymer strands. Here we show that the strength of a network can be greatly amplified by recruiting a larger fraction of strands to bear high tension. We develop approaches to fabricate hydrogels of exceptional strength while maintaining low hysteresis. The strength of the hydrogel increases from ~0.05 MPa for a regular network, to ~1 MPa for a highly entangled network, and further to ~10 MPa for a prestretched interpenetrating network. Similar amplifications of strength are achieved for elastomers. Furthermore, experimental data suggest a scaling relation between strength and strand length. This work provides design principles for creating elastic and strong polymer networks.

## Introduction

A single polymer chain can be pulled, by various techniques, to record force as a function of displacement (*1*). The chain coils entropically before being pulled. As the displacement increases, the force rises gently when the chain uncoils, and steeply when the chain nearly straightens (Fig. 1A). The chain ruptures by breaking a covalent bond. The strength of a polymer network, however, is orders of magnitude below that of a polymer chain (*2-6*). A molecular origin of this large reduction in strength is identified in recent computational simulations (*7-9*). When stretched, a polymer network distributes tension nonuniformly: sparse high-tension paths form and break in succession. Each high-tension path is a sequence of strands (Fig. 1B). The strands off the path remain coiled and carry low tension.

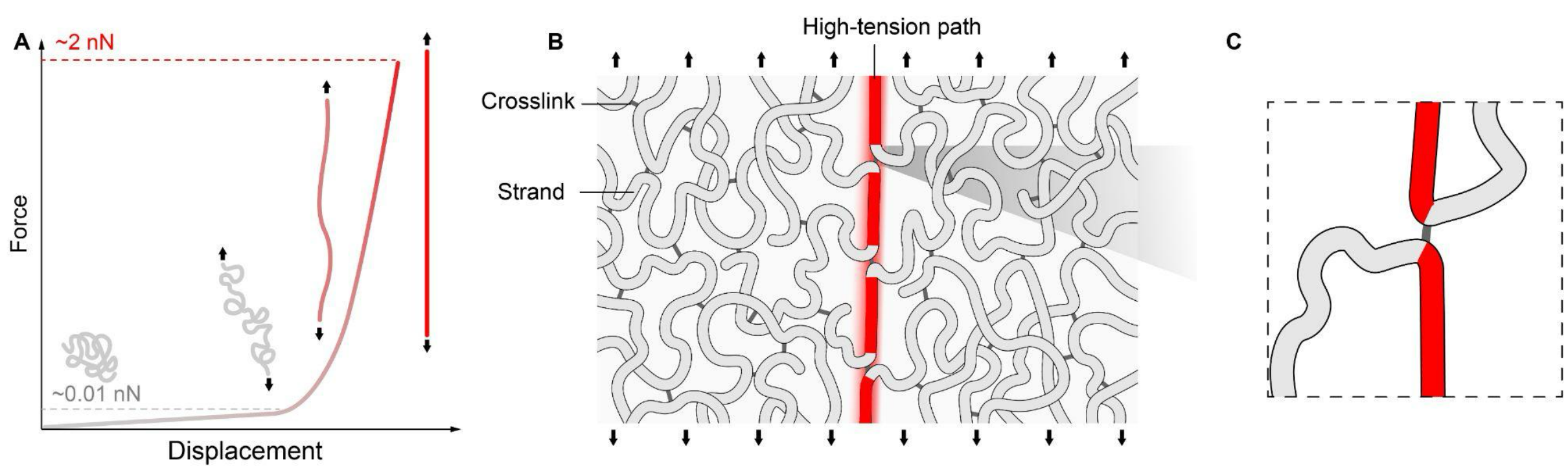

**Fig. 1. Rupture of a polymer chain and a polymer network.** (**A**) Force-displacement curve of a polymer chain. (**B**) In a polymer network of dense crosslinks, strands are short, and a high-tension path is a sequence of strands. (**C**) A crosslink connects four strands, only two of which bear high tension, and the other two do not.

We attribute the concentration of tension to crosslinks. In a polymer network, each crosslink connects four strands by covalent bonds. The noncovalent interactions between strands are much weaker than the covalent bonds, so that high tension can transmit along an entire strand. The strand, however, does not transmit its high tension to all the strands connected to the crosslink. Of the four strands connected by the crosslink, only two bear high tension, and the other two do not (Fig. 1C). It is the crosslinks that concentrate tension.

Here we fabricate polymer networks to achieve high strength by deconcentrating tension. In a polymer network of sparse crosslinks, strands are long and entangled. We hypothesize that such a network forms high-tension cylinders (Fig. 2A). Each high-tension cylinder consists of a volume of long and entangled strands. The diameter of the cylinder in the undeformed state scales with the end-to-end distance of a strand between nearby crosslinks. Because the noncovalent interactions between strands are much weaker than covalent bonds, entanglements readily slip and do not retard the transmission of high tension along the strand. When an entanglement in the cylinder engages four segments, all of them bear high tension (Fig. 2B). Crosslinks concentrate tension and set the diameter of a high-tension cylinder. By contrast, entanglements do not retard the transmission of high tension along each strand, and enable the transmission of high tension from one strand to another within each cylinder. Longer strands promote more uniform tension, engage a larger fraction of strands to bear high tension, and elevate the stress carried by surrounding strands. Our experiments with both gels and elastomers confirm that networks of long strands have high strength. The experimental data also suggest a scaling relation between strength and strand length. Guided by the scaling relation we prepare prestretched interpenetrating networks and demonstrate that they amplify strength further.

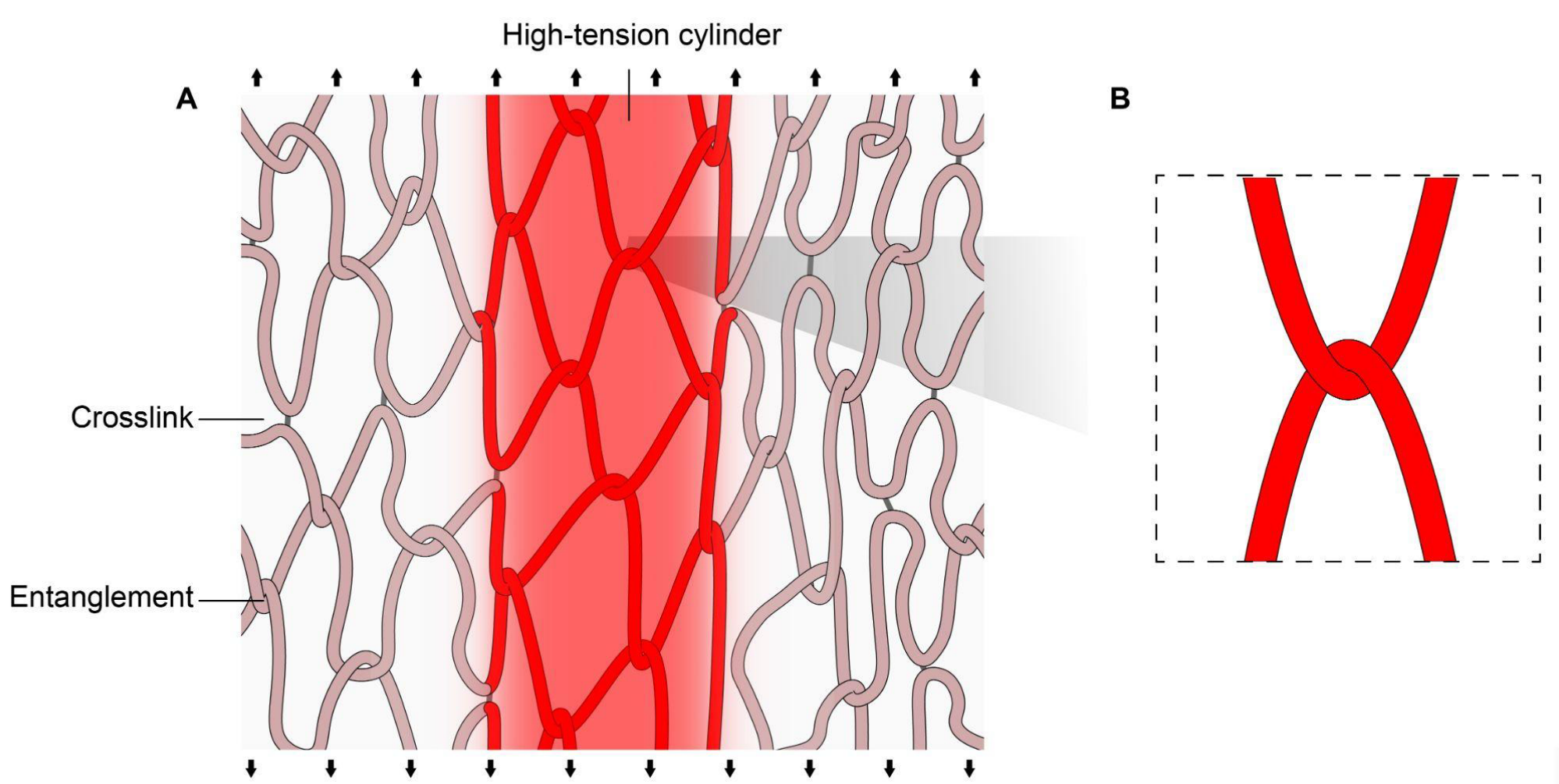

**Fig. 2. High-tension cylinder.** (**A**) In a polymer network of sparse crosslinks, strands are long and entangled, and a high-tension cylinder is a volume of strands. (**B**) In the cylinder, an entanglement engages four segments, all of which bear high tension.

## Results

As illustrations, we demonstrate the strength of hydrogels synthesized under three conditions (Fig. 3A). The strength of the hydrogel increases from ~0.05 MPa for a regular network, to ~1 MPa for a highly entangled network, and further to ~10 MPa for a prestretched interpenetrating network. The hydrogel of a prestretched interpenetrating network can support a weight two orders of magnitude greater than that supported by a regular hydrogel (Fig. 3, B and C; Movies S1 and S2). The influence of network topology on strength will be discussed in detail in the following sections.

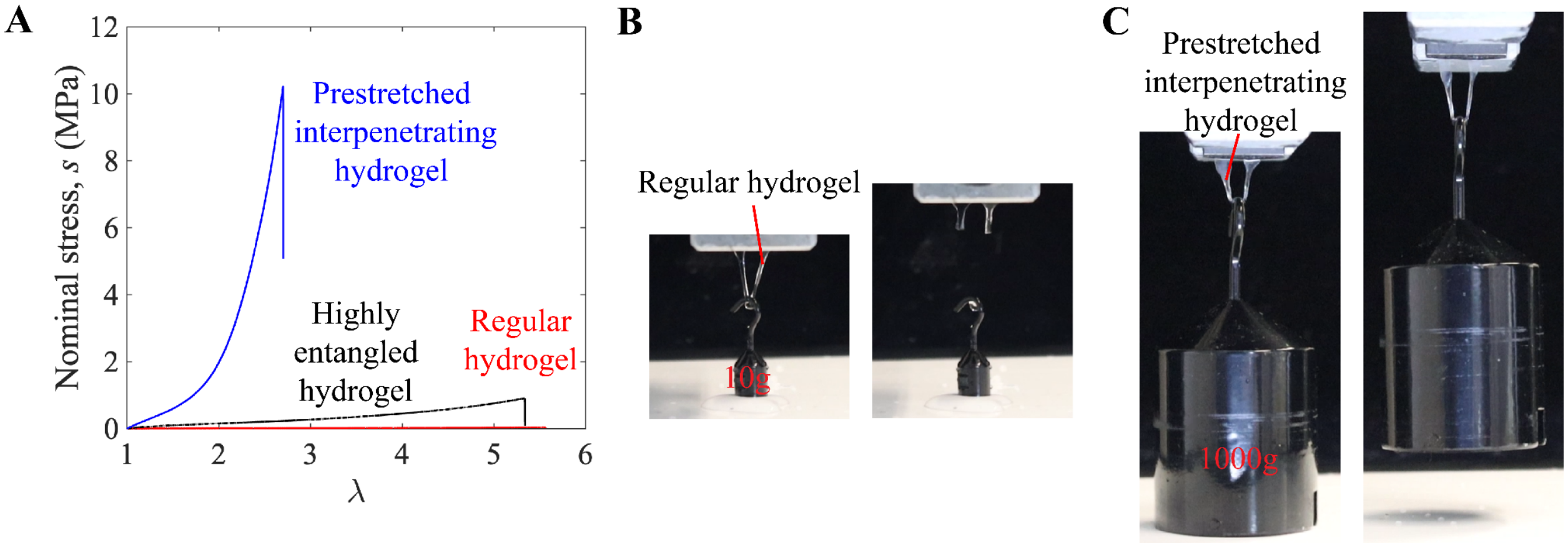

**Fig. 3. Polyacrylamide hydrogels prepared under three conditions**. (**A**) Stress-stretch curves. (**B**) A regular hydrogel, and (**C**) a prestretched interpenetrating hydrogel hanging a weight, respectively.

We test the hypothesis using polyacrylamide (PAAm) hydrogels, prepared by free radical polymerization using precursors composed of four components: monomer, crosslinker, water, and photoinitiator. Let $C$ be the crosslinker-to-monomer molar ratio, and $W$ be the water-to-monomer molar ratio. The photoinitiator-to-crosslinker molar ratio is fixed at 0.4, and the reaction is conducted under an ultraviolet light of low intensity. These conditions ensure that the fraction of dangling ends in the network is negligible, and the network is properly formed (*10-13*). We choose PAAm hydrogels as the model system for two reasons. First, in a fully swollen hydrogel, the polymer chains are screened by water molecules, so that the noncovalent interactions between strands are weak, and their effect on strength is negligible compared to network topology. Second, the densities of crosslinks and entanglements can be independently varied in wide ranges, by changing $C$ and $W$, respectively. In applications, hydrogels are commonly used in contact with water. Thus, the as-prepared hydrogels are submerged in pure water to swell to equilibrium. The fully swollen hydrogels are then used to measure various properties.

The three hydrogels shown in Fig. 3 are prepared as follows. First, we follow the common practice to synthesize hydrogels using precursors with high values of $W = 24$ and $C = 10^{-3}$. The resulting polymer network is densely crosslinked and sparsely entangled. Such materials are referred to as regular hydrogels. Second, we synthesize hydrogels using precursors with low values of $W = 2$ and $C = 10^{-4.5}$. The resulting polymer network is sparsely crosslinked and densely entangled (*10*). Third, to further enhance strength, we prepare a hydrogel of a prestretched interpenetrating network.

The effect of crosslinks on strength is studied by preparing PAAm networks using precursors of a fixed $W = 8$ and various $C$. As noted before, each hydrogel is submerged in pure water and swells to equilibrium before the measurement of various properties. The polymer content $\phi$ is defined as the volume of dry polymer divided by the volume of the fully swollen hydrogel. As expected, the polymer content $\phi$ increases with $C$ (Fig. 4A). Swelling is constrained mostly by entanglements at low $C$, and mostly by crosslinks at high $C$ (*10*).

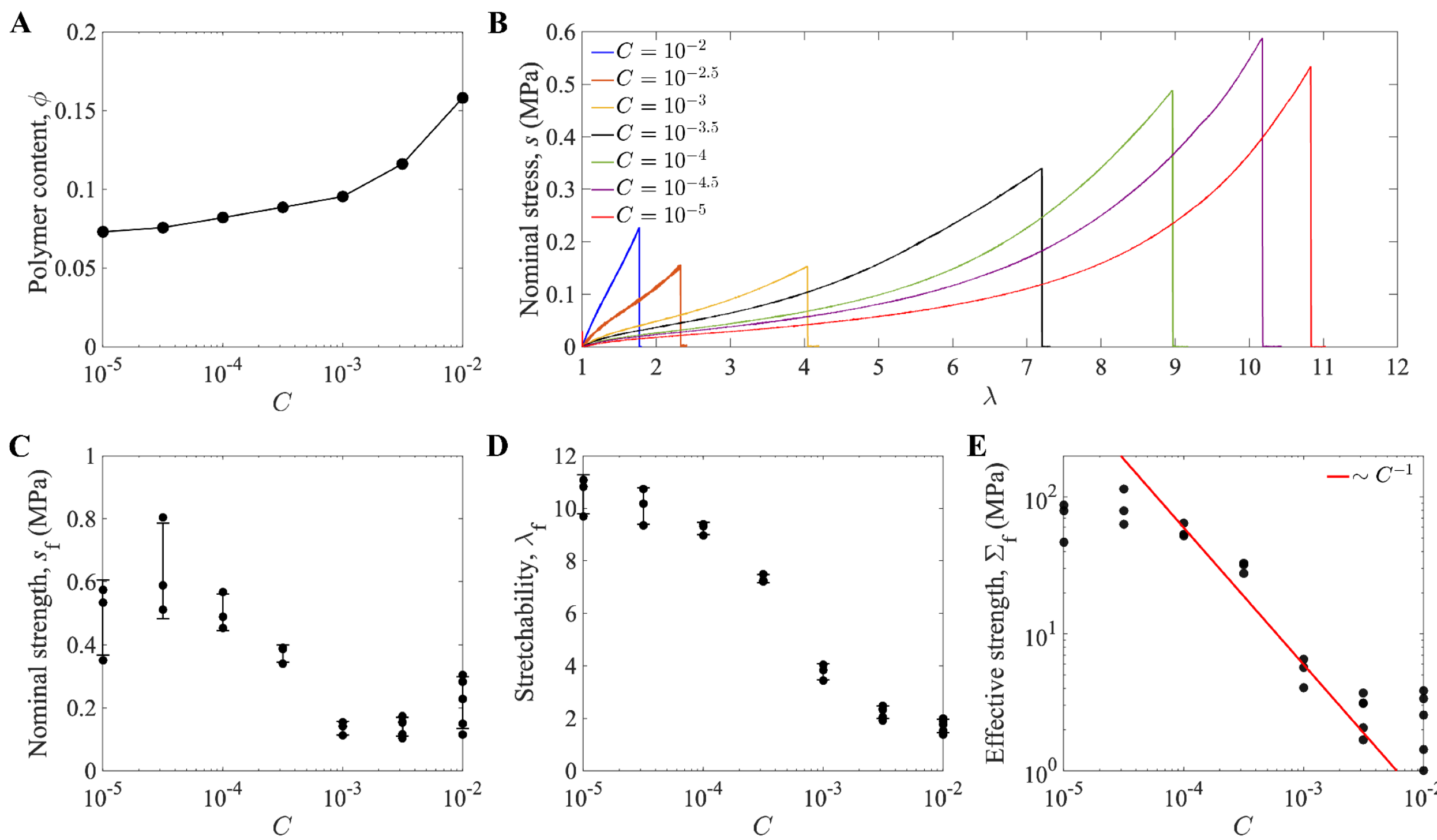

**Fig. 4. Hydrogels prepared using precursors of various crosslinker-to-monomer molar ratios, *C*, and a fixed water-to-monomer molar ratio, *W* = 8.** The as-prepared networks are submerged in pure water to swell to equilibrium before properties are measured as a function of *C*. (**A**) Polymer content $\phi$. (**B**) Stress-stretch curves. (**C**) Nominal strength $s_f$. (**D**) Strechability $\lambda_f$. (**E**) Effective strength $\Sigma_f$.

Stress–stretch curves are measured for samples of various $C$ (Fig. 4B). The nominal stress $s$ is defined as the applied force divided by the cross-sectional area of the undeformed sample. The stretch $\lambda$ is defined as the length of the deformed sample divided by the length of the undeformed sample. The nominal stress and the stretch at rupture define the nominal strength $s_f$ and stretchability $\lambda_f$ (Fig. 4C and D)

The nominal strength $s_f$ is a non-monotonic function of $C$, reaching a maximum at $C \sim 10^{-0.45}$ and a minimum at $C \sim 10^{-2.5}$ (Fig. 4C). When $C < 10^{-4.5}$, the network is poorly formed, and the strength drops. When $C > 10^{-2.5}$, the strength rises again, but scatters widely. These densely crosslinked networks are brittle, so that their strengths are sensitive to flaws. In the rest of this paper, we focus on the intermediate range of $C$, so that the network is well formed, and the strength is flaw-insensitive.

To compare the strength of the polymer network of various stretchability and polymer content, we define the effective stress $\Sigma$ as follows. When stretched, a sample changes its volume negligibly, and is taken to be incompressible. The true stress $\sigma$ is defined as the applied force divided by the cross-sectional area of the deformed sample. The incompressibility relates the true stress to the nominal stress as $\sigma = s\lambda$. In a hydrogel, water molecules do not bear load, so that the effective stress carried by the polymer network itself is the true stress divided by the polymer content, $\Sigma = s\lambda/\phi$ . In an elastomer without solvent, the polymer content is $\phi = 100\%$, and the effective stress coincides with the true stress. In a hydrogel, the polymer content $\phi$ is lower than 100%, and the effective stress exceeds the true stress. The effective strength, $\Sigma_f = s_f \lambda_f / \phi$ , is computed using the measured nominal strength, stretchability, and polymer content.

The effective strength, $\Sigma_f$, varies with the crosslinker-to-monomer molar ratio, $C$ (Fig. 4E). At a low $C = 10^{-4.5}$, both the nominal strength and the stretchability are large, so that the effective strength is high, $\Sigma_f \sim 100$ MPa. At a high $C = 10^{-2.5}$, both the nominal strength and the stretchability are small, so that the effective strength is low, $\Sigma_f \sim 3$ MPa. As noted before, when $C$ is too low ($C < 10^{-4.5}$), hydrogels have poorly formed polymer networks and show low strength. When $C$ is too high ($C > 10^{-2.5}$), hydrogels are brittle, and the strengths are sensitive to flaws and scatter significantly. Once the data in these two extreme regimes are excluded, the effective strength scales with the crosslinker-to-monomer molar ratio as $\Sigma_f \sim C^{-1}$. This scaling indicates that sparse crosslinks enable a larger fraction of strands to approach high tension.

Next, the effect of entanglements on strength is studied by preparing PAAm networks using precursors of various $W$ and a fixed $C = 10^{-3.5}$. When the precursor has a high water-to-monomer molar ratio, $W = 24$, the resulting polymer chains are dilute and sparsely entangled (Fig. 5A). When the precursor has a low water-to-monomer molar ratio, $W = 2$, the resulting polymer chains are concentrated and densely entangled (Fig. 5B). In either case, once the polymer chains are crosslinked into a network, the number of entanglements is invariant when the network is submerged in pure water to swell.

Stress–stretch curves are measured for samples of various $W$ (Fig. S1). As $W$ increases, the polymer content decreases (Fig. 5C), the nominal strength decreases (Fig. 5D), and the stretchability increases (Fig. 5E). Notably, the effective strength is less sensitive to $W$, varying by less than a factor of two (Fig. 5F). Our experimental results support the molecular picture: crosslinks concentrate tension, while entanglements do not. Because noncovalent interactions between strands are much weaker than the covalent bonds along strands, entanglements slip readily under load. That is, an entanglement acts like a pulley, and does not hinder the transmission of high tension along a strand. This "pully effect" between entangled strands is analogous to that in a network containing molecular slide rings (*14, 15*).

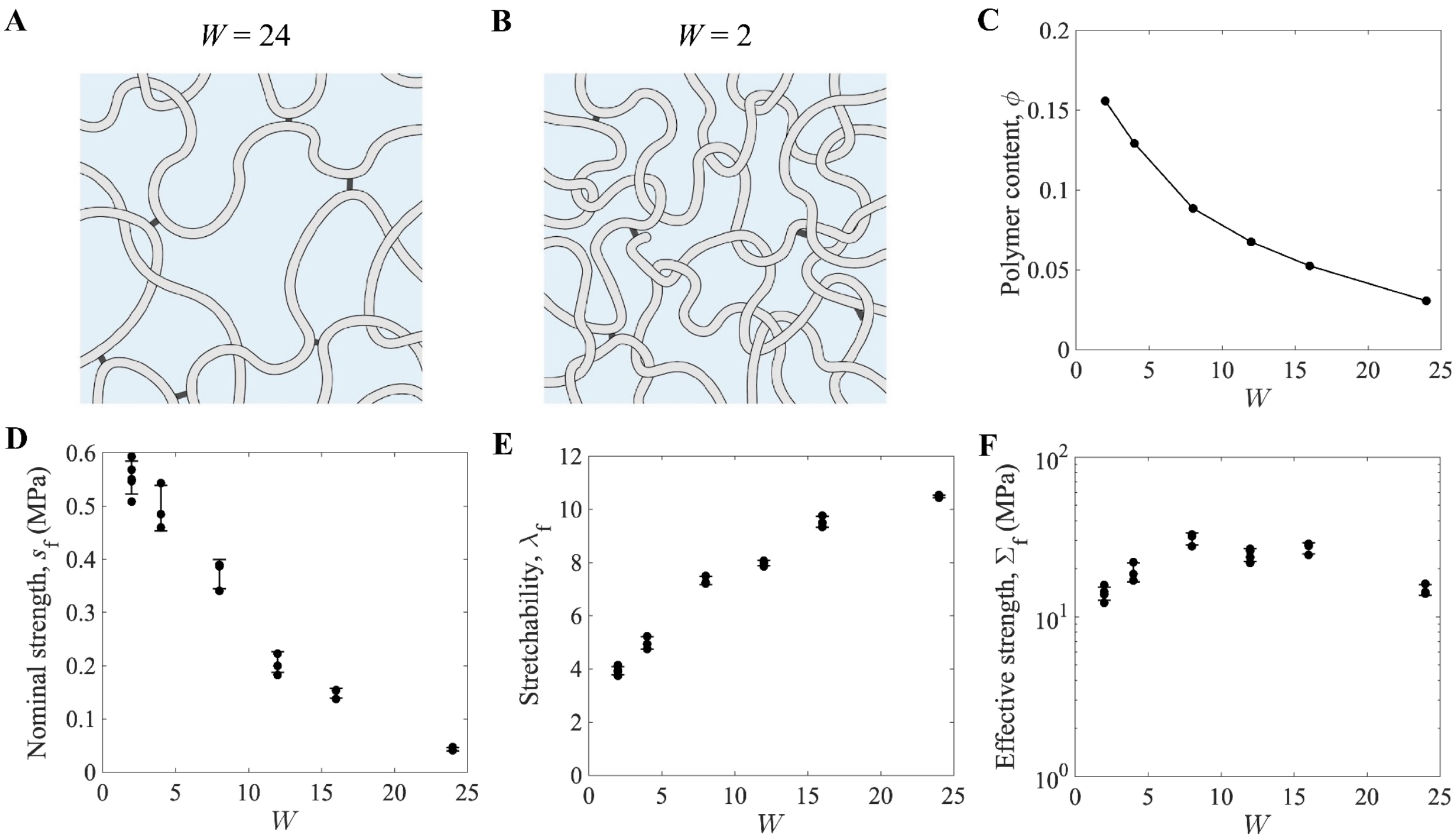

**Fig. 5. Hydrogels prepared with varying water-to-monomer molar ratios, *W*, at a fixed crosslinker-to-monomer molar ratio, *C* = $10^{-3.5}$.** (A) The hydrogel prepared at a high *W* is sparsely entangled. (B) The hydrogel prepared at a low *W* is highly entangled. (C) Polymer content $\phi$. (D) Nominal strength $s_f$. (E) Strechability $\lambda_f$. (F) Effective strength $\Sigma_f$.

We further examine the effect of crosslinks and entanglements on strength by measuring the properties of PAAm hydrogels prepared with various *C* and *W*. These hydrogels exhibit substantial variations in polymer content (Fig. 6A), nominal strength (Fig. 6B), and stretchability (Fig. 6C). By comparison, the effective strength collapses onto a scaling relation of $\Sigma_f \sim C^{-1}$ and shows weak dependence on *W* (Fig. 6D).

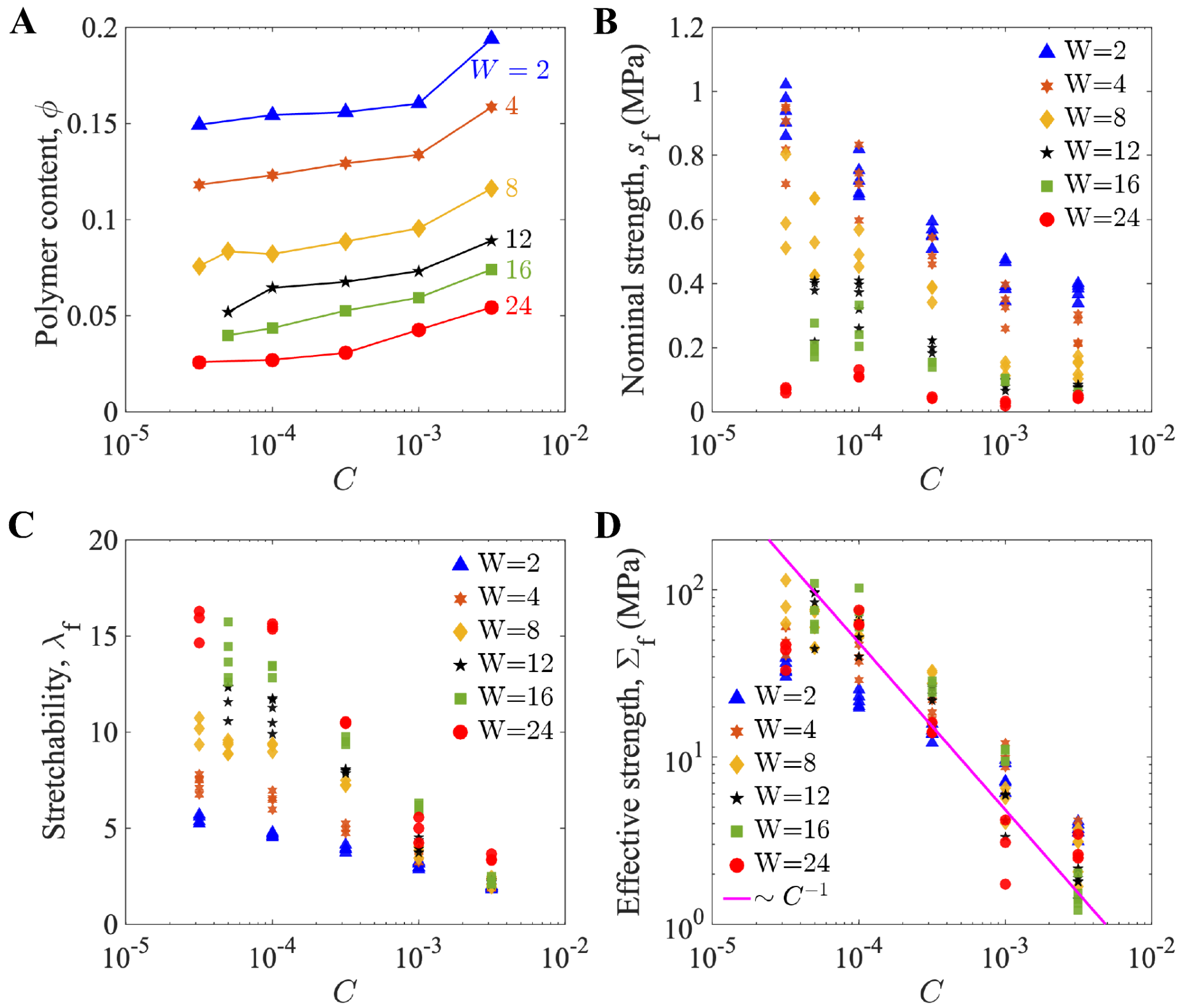

**Fig. 6. Hydrogels prepared using precursors of various C and W.** The hydrogels are then submerged in pure water to swell to equilibrium, before properties are measured. (A) Polymer content $\phi$. (B) Nominal strength $s_f$. (C) Strechability $\lambda_f$. (D) Effective strength $\Sigma_f$.

The effective strength $\Sigma_f$ quantifies the capability of the network to recruit strands to bear high tension, and the nominal strength $s_f$ quantifies the force that can be sustained by a material. The former aids in molecular understanding, and the latter is important in applications. By definition, the two measures of strength are related by $s_f = \phi\Sigma_f/\lambda_f$. The sparse crosslinks set a high effective strength $\Sigma_f$, whereas the entanglements constrain swelling, which leads to a high polymer content $\phi$, and limits the stretch, which leads to a low stretchability $\lambda_f$. The combination of high $\Sigma_f$, high $\phi$, and low $\lambda_f$ leads to a high nominal strength. Among all hydrogels prepared with various $C$ and $W$, the highest nominal strength achieved is $s_f \sim 1$ MPa (Fig. 6B). This high nominal strength is attained by a sample prepared with a precursor with an unusually low value of $C$, which leads to sparse crosslinks, and an unusually low value of $W$, which leads to dense entanglements. By comparison, commonly reported nominal strength for PAAm hydrogels is on the order of 0.05 MPa (*2, 16, 17*).

The above understanding suggests that nominal strength can be amplified by maintaining a high effective strength while increasing polymer content and reducing stretchability. Following this idea, we prepare an interpenetrating network in which strands are long, entangled, and prestretched (Fig. 7A). The long and entangled strands deconcentrate high tension, while the prestretch shifts strand configurations closer to full extension prior to external loading.

We start with an as-prepared single-network (SN) hydrogel of sparse crosslinks and dense entanglements, and then submerge it in the same precursor to swell to equilibrium. The precursor-swollen SN hydrogel is then compressed in the thickness direction into a state of in-plane biaxial prestretch. In the deformed state, upon exposure to ultraviolet light, the precursor polymerizes into a second network, which interpenetrates with the first network. The second network partially locks the deformation of the first network, leading to a prestretched double-network (PDN) hydrogel. The process can repeat itself multiple times. For example, repeating this process twice produces a prestretched triple-network (PTN) hydrogel.

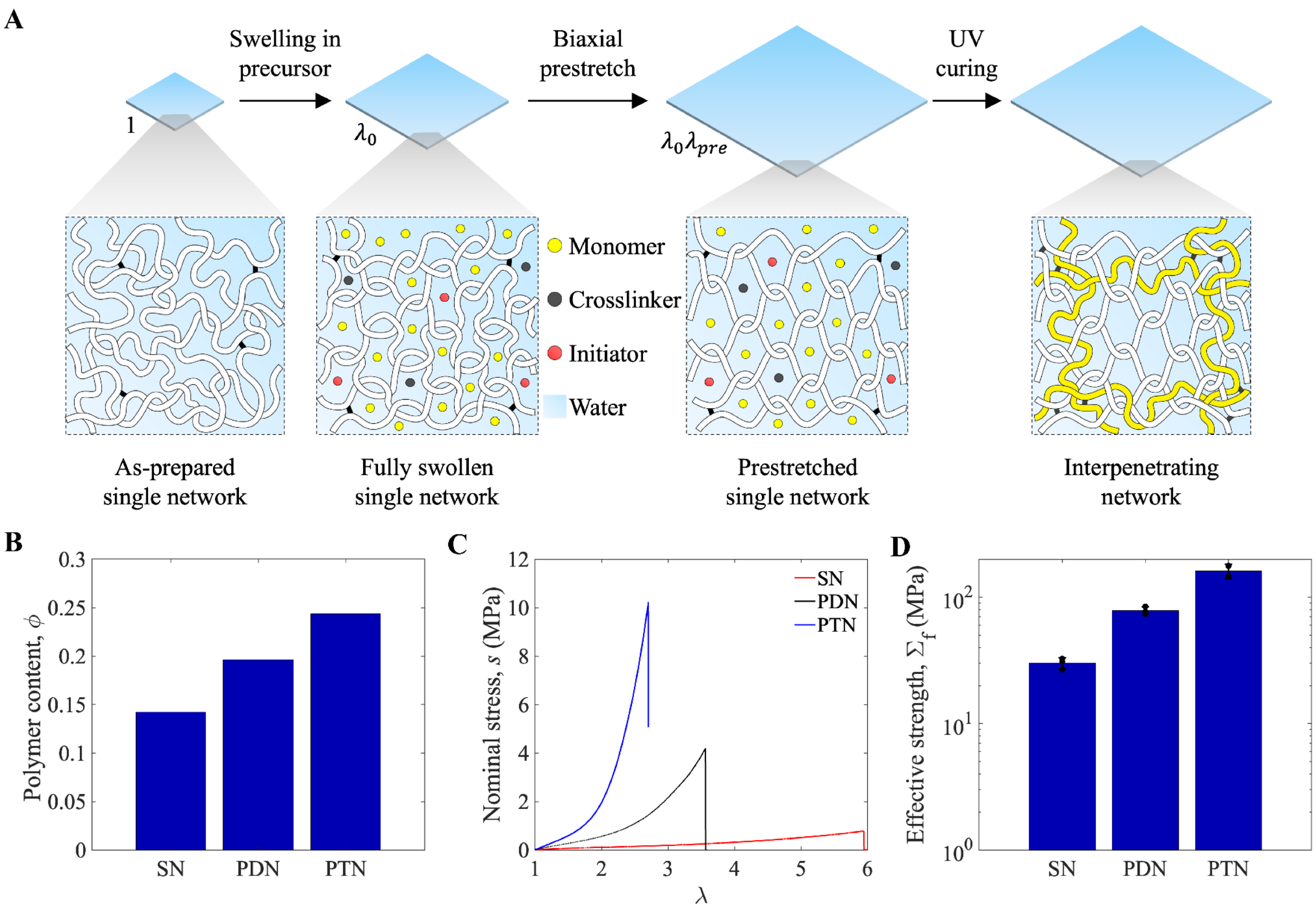

**Fig. 7. Hydrogels of prestretched interpenetrating networks.** (**A**) An as-prepared single-network (SN) hydrogel is fully swollen in a precursor, and then biaxially prestretched, followed by photopolymerization of the second network to form a prestretched double-network (PDN) hydrogel. The process is repeated to form a prestretched triple-network (PTN) hydrogel. All networks are synthesized with the same precursor ($W$ = 3 and $C$ = $5\times10^{-5}$). The resulting hydrogels are fully swollen in water before the test. (**B**) Polymer content. (**C**) stress-stretch curves. (**D**) Effective strength.

All SN, PDN, PTN hydrogels are submerged in pure water to swell to equilibrium. The fully swollen hydrogels are then used to measure polymer contents (Fig. 7B) and stress-stretch curves (Fig. 7C), as well as to calculate the effective strength (Fig. 7D). A PTN hydrogel exhibits an exceptional nominal strength of ~10 MPa, a high polymer content of ~25%, and a moderate stretchability of ~2.8. By contrast, SN hydrogel has a strength of ~0.8 MPa, a moderate polymer content of ~13%, and a high stretchability of ~6.

The high nominal stress is contributed by the two processes during the preparation of interpenetrating networks, namely swelling and mechanical prestretch. First, each round of precursor swelling leads to an isotropic prestretch of the pre-existing polymer network. Second, the mechanical prestretch further straightens the pre-existing network strands. The prestretch effect limits the stretchability (Fig. S2). The reduced stretchability does not result from the brittleness or premature fracture. Rather, it is because more strands in the network reach their full extension at a smaller stretch. This interpretation is supported by the pronounced strain-stiffening of the PTN hydrogel before rupture, consistent with the single-chain mechanics that the chain stiffens as it approaches its full extension.

Moreover, the effective strength of the PTN is about 115 MPa, ~2.6 times that of SN about 45 MPa, and still falls in the scaling plane of the effective strength range at $C = 5\times10^{-5}$ (Fig. S3). The enhancement in effective strength of PTN compared to the SN likely results from more efficient stress deconcentration across neighboring strands by interpenetrating structure, enabling more strands to bear high tension.

The high effective strength set by sparse crosslinks, together with the high polymer content and limited stretchability set by entanglements and the prestretch, lead to the high nominal strength of PTN hydrogel. To the best of our knowledge, the strength of PTN hydrogel is the highest among the polyacrylamide hydrogels.

In addition to the high strength, the PTN hydrogel remains highly elastic, with a low hysteresis of 8.4% (Fig. S4). Here, hysteresis is defined by the ratio of two areas: the area enclosed by the loading-unloading curves, and the area underneath the loading curve, measured up to the stretch of two. The low hysteresis results from two aspects. First, the noncovalent interactions between strands are weak. Second, strands are long and remain intact at a moderate stretch. In the plane of hysteresis and nominal strength, the PTN hydrogel shows both high strength and low hysteresis, outperforming existing PAAm-based hydrogels (Fig. 8). Also note that high nominal strength has been achieved by some polymer networks with significant noncovalent interactions between strands. Examples include natural rubber and PEG hydrogels that undergo strain induced crystallization, PVA hydrogels and polyurethane that contain hard domains, and glassy gels strengthened by solvent-mediated noncovalent crosslinking (*18-23*). These noncovalent interactions usually increase hysteresis. And how noncovalent interactions enhance strength requires further study.

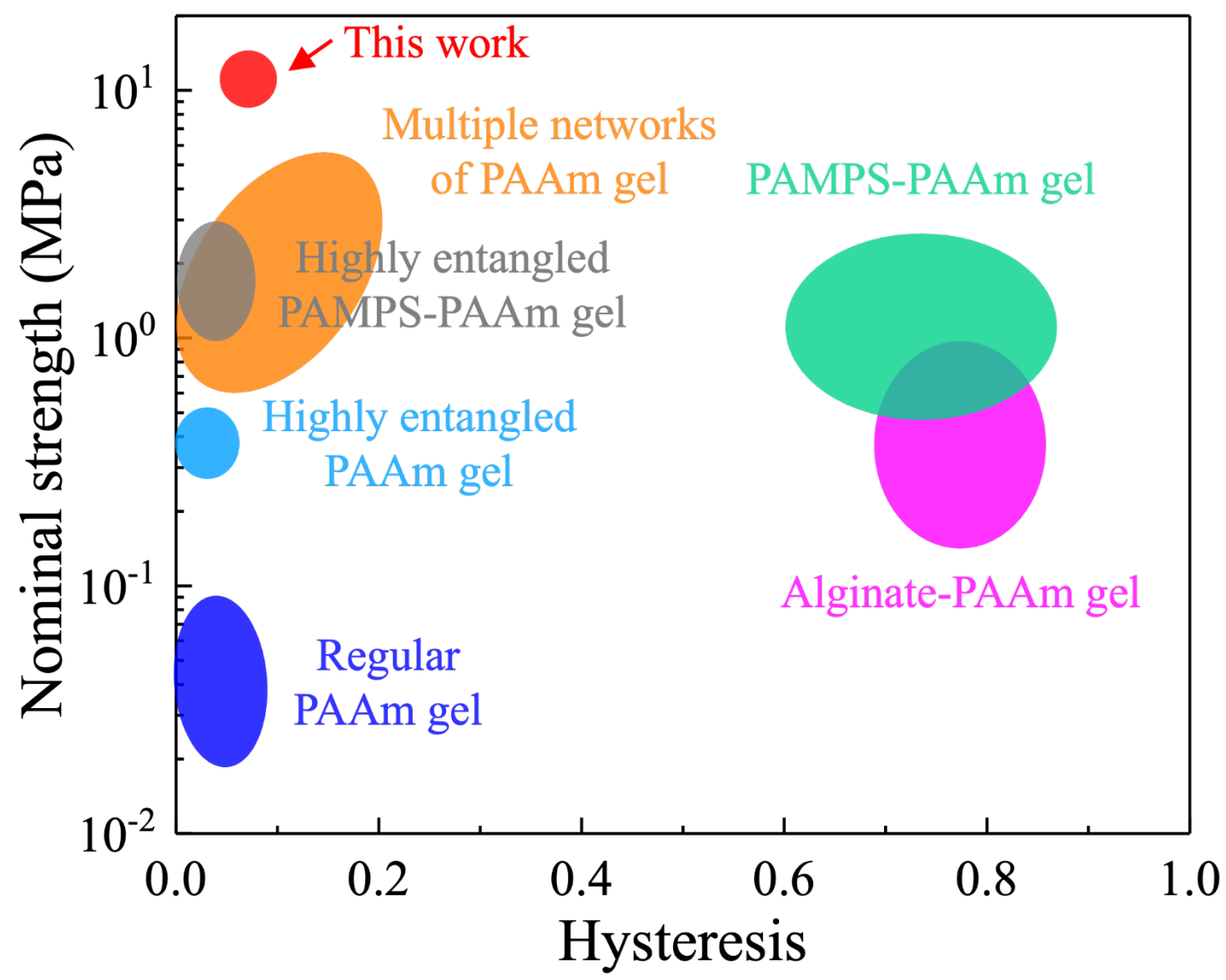

**Fig. 8.** Comparison of various polyacrylamide-based hydrogels on the plane of nominal strength and hysteresis. (*2, 10, 24-28*).

To demonstrate the generality of our design principles, we prepare a model elastomer, poly(ethyl acrylate) (PEA), by free radical polymerization using precursors composed of three components: monomer, crosslinker, and photoinitiator. Because no extra solvents are used, the entanglement density is invariant. The nominal stress–stretch curves are measured for PEA with different crosslinker-to-monomer molar ratios *C* (Fig. 9A). The nominal strength increases from ~1 MPa for a highly crosslinked network, to ~4 MPa for a sparse crosslinked network (Fig. 9B). The effective strength also roughly follows the $\Sigma_f \sim C^{-1}$ scaling relation (Fig. 9C). We further amplify the nominal strength of the sparsely crosslinked network by making a prestretched double-network (PDN) PEA (Fig. 9D). As the ratio of mechanical prestretch increases, more strands approach their full extension prior to external loading, so that the nominal strength further enhances to ~25 MP (Fig. 9E). In addition to this exceptional strength, the PDN network maintains a low hysteresis of ~5% (Fig. 9F).

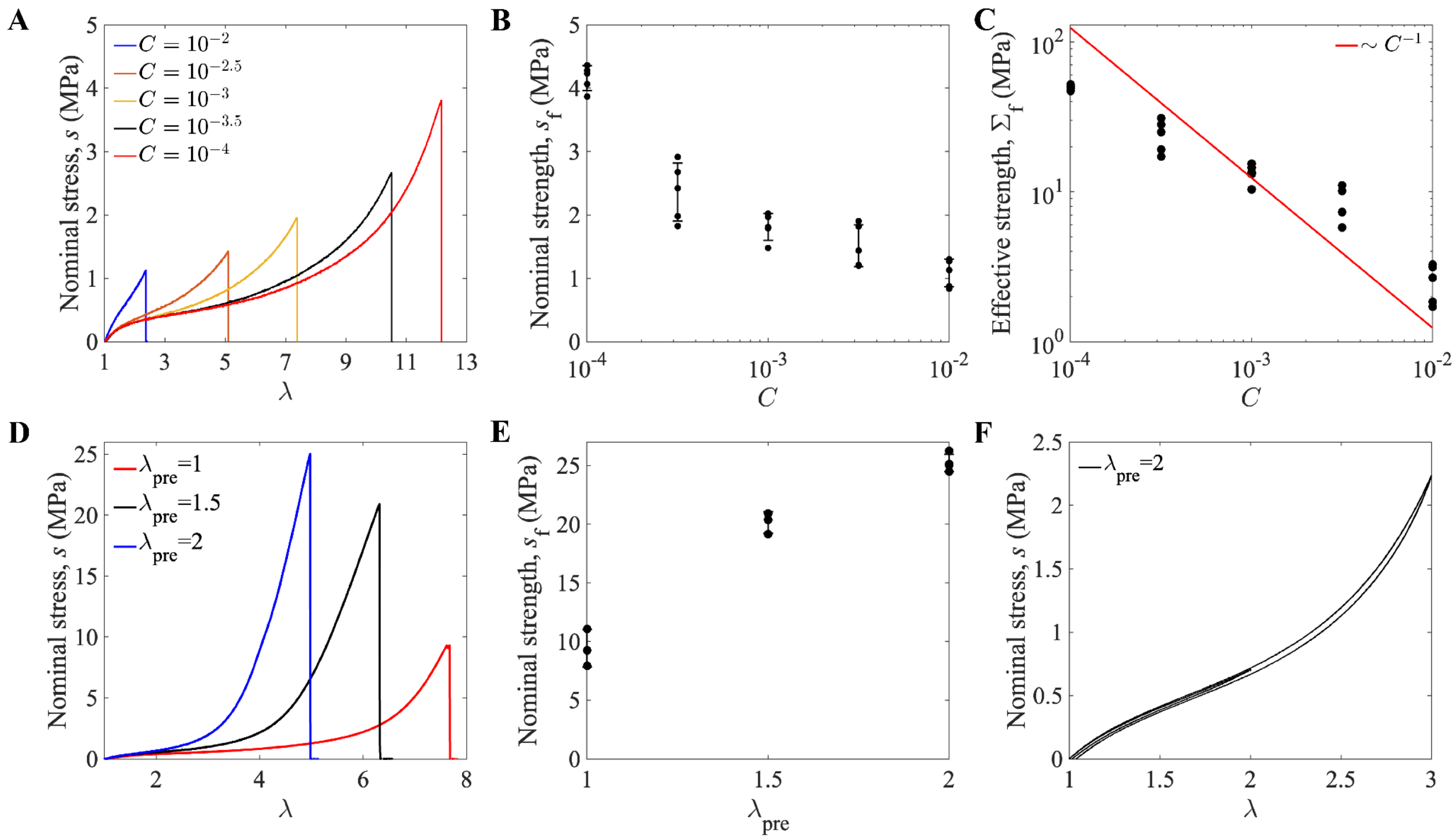

**Fig. 9. Mechanical properties of elastomers.** (**A**) Stress–stretch curves of single-network (SN) poly(ethyl acrylate) (PEA) elastomers prepared with varying *C*. (**B**) Nominal strength $s_f$ as a function of *C* for SN elastomers. (**C**) Effective strength of SN. The effective strength of SN elastomers as a function of *C* is fitted to the $\Sigma_f \sim C^{-1}$ scaling. (**D**) Stress–stretch curves of double-network (DN) elastomers prepared with different prestretch ratios $\lambda_{pre}$. (**E**) Nominal strength $s_f$ as a function of $\lambda_{pre}$ for DN elastomers. (**F**) The loading-unloading curve of the DN elastomer prepared with $\lambda_{pre}$=2. The elastomer has a low hysteresis of 4.3% for loading up to $\lambda = 2$ and 6.2% for $\lambda = 3$.

## Discussion

In summary, we have shown that the strength of a network can be greatly amplified by recruiting a larger fraction of strands to bear high tension. We fabricate hydrogels of exceptional strength by several approaches, including forming polymer networks with long and entangled strands, interpenetrating networks, and prestretched networks. The strength is ~0.05 MPa for a regular network, ~1 MPa for a highly entangled network, and ~10 MPa for a prestretched interpenetrating network. Similar amplifications of strength are achieved for elastomers. Furthermore, experimental data suggest a scaling relation between strength and strand length. This work provides design principles for creating elastic and strong polymer networks.

Our experimental data indicate an approximate scaling of $\Sigma_f \sim C^{-1}$. Several observations provide some understanding. As noted before, in a polymer network of sparse crosslinks, strands form high-tension cylinders (Fig. 2A). Each high-tension cylinder consists of a volume of long and entangled strands. Crosslinks concentrate tension and set the diameter of a high-tension cylinder. In the undeformed state, the diameter of such a cylinder scales with the end-to-end distance of a strand between nearby crosslinks, which follows $C^{-0.5}$ for the random walk model. Therefore, the cross-sectional area of the high-tension cylinder in the undeformed state scales as $C^{-1}$. Because the noncovalent interactions between strands are much weaker than covalent bonds, entanglements readily slip and do not retard the transmission of high tension along the strand. When an entanglement in the cylinder engages four segments, all of them bear high tension (Fig. 2B). These considerations are consistent with the experimental findings that high effective strength is achieved at a low crosslink density, and is not sensitive to either entanglement density or prestretch.

Polymer networks of sparse crosslinks, the highest effective strength achieved is $\Sigma_f \sim 100$ MPa, which is, however, still much smaller than the strength of covalent bonds of ~10 GPa. This large reduction in strength indicates that within the polymer network the high-tension cylinders are sparse. About 1% of strands achieve the covalent bond strength, while the strands off high-tension cylinders carry much lower tension. How to recruit more strands to bear high tension remains an open question and may require new network architectures.

## Materials and Methods

### Preparation of single-network (SN) PAAm hydrogel

Acrylamide (AAm, A8887), N,N′-methylenebisacrylamide (MBAA, M7279), and 2-hydroxy-4′-(2-hydroxyethoxy)-2-methylpropiophenone (Irgacure 2959, 410896) were used as the monomer, crosslinker, and photoinitiator, respectively. All chemicals were purchased from Sigma-Aldrich and used as received. Deionized water was used as the solvent and obtained from Poland Spring. The monomer, crosslinker, and initiator solutions were mixed with water in a conical tube to obtain precursors with different water-to-monomer ($W$), crosslinker-to-monomer ($C$), and initiator-to-monomer ($I$) molar ratios. The initiator-to-crosslinker molar ratio is fixed at $I/C = 0.4$. A mold was prepared by pressing a silicone spacer (thickness 0.8 mm) onto a glass plate. The precursor solution was poured into the mold and covered with a second glass plate. The two plates were clamped together using binder clips. The assembled mold was placed inside a nitrogen-filled plastic bag and irradiated for 3hours using a UV lamp (365 nm, ~5 mW cm$^{-2}$). After curing, the hydrogel sample was submerged in water for at least 24 hours to swell to equilibrium.

### Preparation of double-network (DN) PAAm hydrogel

The first network was prepared from a precursor solution with a water-to-monomer molar ratio $W = 3$ and a crosslinker-to-monomer molar ratio $C = 5\times10^{-5}$. The resulting SN hydrogel was then soaked in the same precursor for 24 hours to fully swell. The swollen hydrogel was placed between two glass plates, with the edges sealed using parafilm. The plates were compressed using eight C-shaped clamps (McMaster-Carr, 5133A13) to achieve various prestretch $\lambda_{pre}$. The PDN hydrogel in Fig.7 is synthesized with $\lambda_{pre} = 2$. The assembled mold was placed in a nitrogen-filled plastic bag and irradiated for 3 hours using a UV lamp (365 nm, ~5 mW cm$^{-2}$). After curing, the DN hydrogel was submerged in water for at least 24 hours to swell to equilibrium.

### Preparation of triple-network (TN) PAAm hydrogel

The TN hydrogel was prepared by soaking a DN hydrogel in a precursor solution with $W = 3$ and $C = 5\times10^{-5}$, followed by photopolymerization. TN-free hydrogels were obtained without applying any prestretch during the polymerization of both the second and third network. To prepare PTN hydrogels, a DN hydrogel (synthesized with $\lambda_{pre} = 2$) was soaked in the same precursor solution and prestretched to $\lambda_{pre} = 1.5$ during the polymerization of the third network.

### Preparation of single-network (SN) PEA

Ethyl acrylate (EA), tricyclo[5.2.1.0$^{2,6}$]decanedimethanol diacrylate (TDDA), and 2-hydroxy-2-methylpropiophenone (Irgacure 1173) were used as the monomer, crosslinker, and photoinitiator, respectively. All reagents were purchased from Sigma-Aldrich and used as received. Solutions of the initiator and crosslinker were mixed with the monomer in a conical tube. The initiator-to-crosslinker molar ratio was fixed at 0.4, while the crosslinker-to-monomer molar ratio $C$ was varied. A mold was prepared by pressing a silicone spacer (thickness 0.8 mm) onto a glass plate. The precursor solution was poured into the mold and covered with a second glass plate. The two plates were clamped together using binder clips. The assembled mold was placed inside a nitrogen-filled plastic bag and irradiated overnight using a UV lamp (365 nm, ~ 5 mW $cm^{-2}$). After curing, the sample was removed from the mold and left in a fume hood for 24 hours to allow evaporation of residual monomers.

**Preparation of double-network (DN) PEA**

The SN network was prepared from a precursor solution with a crosslinker-to-monomer molar ratio $C$=$10^{-4}$. The elastomer was then soaked in a second precursor solution of the same composition for 24 hours to allow full swelling. The swollen elastomer was placed between two glass plates, and the edges were sealed using parafilm. The plates were compressed using eight C-shaped clamps to achieve various prestretch $\lambda_{pre}$. The assembled mold was enclosed in a nitrogen-filled plastic bag and irradiated for 12 hours using a UV lamp (365 nm, ~ 5 mW $cm^{-2}$).

**Tensile tests**

Samples were cut into a dumbbell shape using a die cutter (Ace Steel Rule Dies, ISO 527-2-5B). All mechanical tests were conducted with Instron 5966 universal testing machine equipped with a 100 N load cell. The gauge length and width of the sample were 20 mm and 2 mm, respectively. All tensile tests were performed at a constant stretch rate of $0.05\ s^{-1}$. A humidifier was used to provide vapor to hydrogel samples during the tensile tests to prevent water loss.

**Polymer content measurement**

The polymer content $\phi$ was calculated as the volume of dry polymer divided by the volume of the fully swollen hydrogel. $\phi$ was calculated as $\phi = 1/[(m/m_{dry} - 1)(\rho_A/\rho_w) + 1]$, where $m$ is the mass of the fully swollen hydrogel, $m_{dry}$ is the dry mass, $\rho_A$ = 1.13 g/$cm^3$ is the density of acrylamide, and $\rho_w$ = 1 g/$cm^3$ is the density of water.

## Acknowledgments

**Funding**: This work was supported by the National Science Foundation under MRSEC (DMR-2011754).

**Author contributions**: Y.Z. conceived of the project. Y.Z., C.L., Z.C., and Z.S. designed the study and interpreted the results. Y.Z. conducted the experiments. Y.Z., C.L., Z.C., X.B., and Z.S. wrote the manuscript. Z.S. supervised the research. All authors discussed the result and commented on the manuscript.

**Competing interests:** The authors declare that they have no competing interests.